\documentclass[sigconf]{acmart}
\usepackage{graphicx}
\usepackage{enumitem}
\usepackage{tabularx}
\usepackage{ulem}
\usepackage{float}
\usepackage{soul}
\usepackage{xcolor}
\usepackage{titlesec}
\usepackage{subcaption} 
\usepackage{multirow}
\usepackage{booktabs}
\usepackage{array}
\AtBeginDocument{%
  \providecommand\BibTeX{{%
    \normalfont B\kern-0.5em{\scshape i\kern-0.25em b}\kern-0.8em\TeX}}}

\setcopyright{acmlicensed}
\copyrightyear{2024}
\acmYear{2024}
\acmDOI{XXXXXXX.XXXXXXX}

\acmConference[ICSE 2024]{46th International Conference on Software Engineering}{April 2024}{Lisbon, Portugal}
\acmISBN{978-1-4503-XXXX-X/18/06}

\begin{document}

\title{Unit Test Generation using Generative AI : 
A Comparative Performance Analysis of Autogeneration Tools
}

\author{Shreya Bhatia\textsuperscript{$ \dag  $}}
\email{shreya20542@iiitd.ac.in}
\affiliation{%
  \institution{IIIT Delhi}
  \state{Delhi}
  \country{India}
}

\author{Tarushi Gandhi\textsuperscript{$ \dag  $}}
\email{tarushi20579@iiitd.ac.in}
\affiliation{%
  \institution{IIIT Delhi}
  \state{Delhi}
  \country{India}
}
\author{Dhruv Kumar}
\email{dhruv.kumar@iiitd.ac.in}
\affiliation{%
  \institution{IIIT Delhi}
  \state{Delhi}
  \country{India}
}
\author{Pankaj Jalote}
\email{jalote@iiitd.ac.in}
\affiliation{%
  \institution{IIIT Delhi}
  \state{Delhi}
  \country{India}
}

\titlenote{\textsuperscript{$ \dag  $}These authors contributed equally.}

\renewcommand{\shortauthors}{Bhatia, Gandhi et al.}

\begin{abstract}
Generating unit tests is a crucial task in software development, demanding substantial time and effort from programmers. The advent of Large Language Models (LLMs) introduces a novel avenue for unit test script generation. This research aims to experimentally investigate the effectiveness of LLMs, specifically exemplified by ChatGPT, for generating unit test scripts for Python programs, and how the generated test cases compare with those generated by an existing unit test generator (Pynguin). For experiments, we consider three types of code units: 1) Procedural scripts, 2) Function-based modular code, and 3)  Class-based code. The generated test cases are evaluated based on criteria such as coverage, correctness, and readability. 
Our results show that ChatGPT’s performance is comparable
with Pynguin in terms of coverage, though for some cases its performance is 
 superior to Pynguin. We also find that about a third
 of assertions generated by ChatGPT for some categories were incorrect. 
Our results also show that there
is minimal overlap in missed statements between ChatGPT and
Pynguin, thus, suggesting that a combination of both tools may
enhance unit test generation performance. 
Finally, in our experiments, prompt engineering improved ChatGPT’s performance, achieving a much higher coverage.
\end{abstract}

\maketitle

\section{Introduction}\label{sec:intro}
Unit testing is an integral part of software development as it helps catch errors early in the development process. Creating and maintaining effective unit tests manually is a notably laborious and time-consuming task. To address the difficulties inherent in manual test creation, various methodologies for automating the unit test generation process have been proposed by researchers. Common approaches in this field include search-based \cite{Fraser2012SBST3, Andrews2011SBST2, Baresi2010SBST4, derakhshanfar2022SBST1, Tonella2004SBST5}, constraint-based \cite{Ma2015constraint1, Sakti2015constraint2}, or random-based \cite{Csallner2004random1, Pacheo2007random2} techniques, all aiming to generate a suite of unit tests with the primary goal of enhancing coverage in the targeted software.
However, when compared to tests that are manually created, automated tests produced by these techniques could be less readable and comprehensible \cite{Gargari2021SBSTchallenges1, Harman2015SBSTchallenges2}. This shortcoming makes it difficult for testers with little experience to learn and hinders the adoption of these strategies, especially for beginners. As a result, developers might be hesitant to include these automated tests straight into their workflows. To address these concerns, recent efforts have explored the use of advanced deep learning (DL) techniques, particularly large language models (LLMs), for unit test generation \cite{Dinella_2022toga, lahiri2023ticoder, lemieux2023codamosa}. 

Advancements in Large Language Models (LLMs), exemplified by OpenAI's ChatGPT (GPT-3.5), showcase enhanced capabilities in routine tasks like question answering, translation, and text/code generation, rivaling human-like understanding. Unlike other LLMs such as BART\cite{lewis2019bart} and BERT\cite{devlin2019bert}, ChatGPT \cite{brown2020chatgpt} incorporates reinforcement learning from human feedback (RLHF) \cite{rlhf2017christiano} and a larger model scale, improving generalization and alignment with human intention. Widely utilized in daily activities, ChatGPT is crucial for tasks like text generation, language translation, and automated customer support. Beyond daily use, large language models are increasingly applied in software engineering tasks, including code generation and summarization\cite{wang2023codeGeneration, ahmed2023codeSummarization}. These models can also facilitate the generation of unit test cases, streamlining software validation processes\cite{schäfer2023softwareValidation, yu2023softwareValidation, lemieux2023codamosa}.

This paper seeks to explore the advantages and drawbacks of test suites produced by Large Language Models. We focus particularly on ChatGPT as a representative of the LLMs. Furthermore, we want to explore the potential synergy of integrating existing unit test generators such as Pynguin \cite{Lukasczyk_2022Pynguin} with Large Language Model (LLM)-based approaches to enhance overall performance.  

We evaluate the quality of unit tests generated by ChatGPT compared to Pynguin. Based on the code structure, we classify a sample Python code into 3 categories: 1) Procedural scripts, where code does not have classes or functions. 
2) Function-based modular code is where there are clear definitions of functions which are standalone and act like independent units of code. and 3) Class-based modular code, which is structured around classes and objects, as the primary units of organisation. 

We curated a dataset comprising 60 Python projects, categorizing them into 20 projects per category, each with a complete executable environment. In our study, we focus on a designated core module from each project, ranging between 100-300 lines of code, selected based on factors such as cyclomatic complexity, function count, and file interdependency. We then generate unit tests for the selected modules by prompting them as input to ChatGPT and compare them with unit tests generated by Pynguin. We aim to address the following research questions:
\begin{itemize}
    \item \textbf{RQ1 (Comparative Performance):} How does ChatGPT compare with Pynguin in generating unit tests?
    \item \textbf{RQ2 (Performance Saturation and Iterative Improvement):} How does the effectiveness of test cases generated by ChatGPT improve/change over multiple iterations of prompting?
    \item \textbf{RQ3 (Quality Assessment):} How correct are the assertions generated by ChatGPT, and what percentage of assertions align with the intended functionality of the code?
    \item \textbf{RQ4 (Combining Tools for Improved Performance):} Can a combination of ChatGPT and Pynguin enhance the overall performance of unit test generation, in terms of coverage and effectiveness?
   
\end{itemize}

This paper validates the findings from existing work 
\cite{lemieux2023codamosa} which is very important in the rapidly evolving landscape of LLMs and additionally explores research questions not covered in the existing work.

The rest of the paper is organized as follows: We explain the methodology in \S \ref{sec:method} followed by results in \S \ref{sec:results}. We discuss related work in \S \ref{sec:rw} and conclude in \S \ref{sec:conclusion}. 
\section{Methodology}\label{sec:method}
In this section, we discuss the systematic approach undertaken to compare the performance of various unit test generation tools, including the selection of code samples, their categorisation, tool choices, and evaluation metrics. 

\subsection{Categorisation Based on Code Structure}
The three delineated categories served to capture varying levels of code organisation:

\textbf{Category 1 (Procedural Scripts):} 
Code samples are characterised by procedural scripts lacking defined classes and functions. Many Python programs are scripts of this type.

\textbf{Category 2 (Function-based modular code):} 
Code samples with definitions of standalone functions that
act as independent units of code. Such organization of code offers
limited encapsulation with potential reliance on global variables
and no inherent hiding or protection of the data within functions.

\textbf{Category 3 (Class-based modular code):} 
Code samples containing defined classes and methods. 
Most of the larger code samples collected belonged to
this category.

\subsection{Data Collection}
To conduct a comprehensive evaluation of unit test generation tools, we began by gathering a dataset of Python code samples, encompassing a diverse range of projects. Pynguin has a limitation that it does not work well with Python programs that make use of native code, such as Numpy \cite{Lukasczyk_2022Pynguin}\cite{lukasczyk_empirical_2023testGenerationPython}. For our comparative analysis, we had to make use of Python projects that did not have a dependency on such libraries. 

Initially, we selected about 60 Python projects from open source Github repositories\footnote {The selected projects from all the GitHub repositories can be found here: https://github.com/Rey-2001/LLM-nirvana} , ensuring a diverse and balanced representation while being mindful of Pynguin’s limitation. 
Later we added an additional set of 49 Python Projects from the benchmark data used by Lukasczyk et al. \cite{Lukasczyk_2022Pynguin}.
In total, we have 109 Python projects.

Each project contained multiple files, having import dependencies on one another. For the purpose of our test-generation experiments, we decided to select one core module (i.e. one file) from each project, that we would pass as the prompt input to ChatGPT.
We initially limited our selection of core modules to a size of 0-100  lines of code (LOC), which we will call ‘Small Code Samples’. 
We then expanded the scope to consider core module files of sizes ranging from 100 to 300 LOC, we will refer to them as ‘Large Code Samples’.

For selecting one core module from each project, we narrowed down our selection to files lying in the required LOC range. From these files, we selected the ones which had the highest McCabe complexity \cite{Mccabe1976} (also known as cyclomatic complexity). 
For files with similar complexity, we further looked at the number of functions, the richness of logic, and how frequently they were being referenced or imported by other files. Files having high complexity, more function count, and a higher number of import dependencies were chosen as core modules. The goal was to select modules that can be understood on their own without providing prior context, but will also not be too trivial for testing. 

We were able to collect a total of 60 core modules under 'Small Code samples', with 20 modules belonging to each of the three categories. And under the 'Large Code Samples', we collected 49 core modules, with 20 modules in Category 2, 23 modules in Category 3 and 6 in Category 1 (very few fell in this category, as most of the projects follow a modular programming approach for easier maintenance).

\subsection{Unit Test Generation Tools}
We explored the potential of recent tools, Ticoder and Codamosa; however, due to their unavailability for direct experimentation, we opted for Pynguin, a proficient Python unit test generation tool, to further carry out a comprehensive comparison with ChatGPT. We engaged ChatGPT and Pynguin to generate unit test cases for each of the identified core modules. 

\subsection{Prompt Design for ChatGPT}
We design our prompt by using clear and descriptive words to capture the intent behind our query, based on the widely acknowledged experience of utilizing ChatGPT\cite{openai_doc}\cite{prompt_design}. 

\begin{figure}
  \centering
  \includegraphics[width=0.45\textwidth]{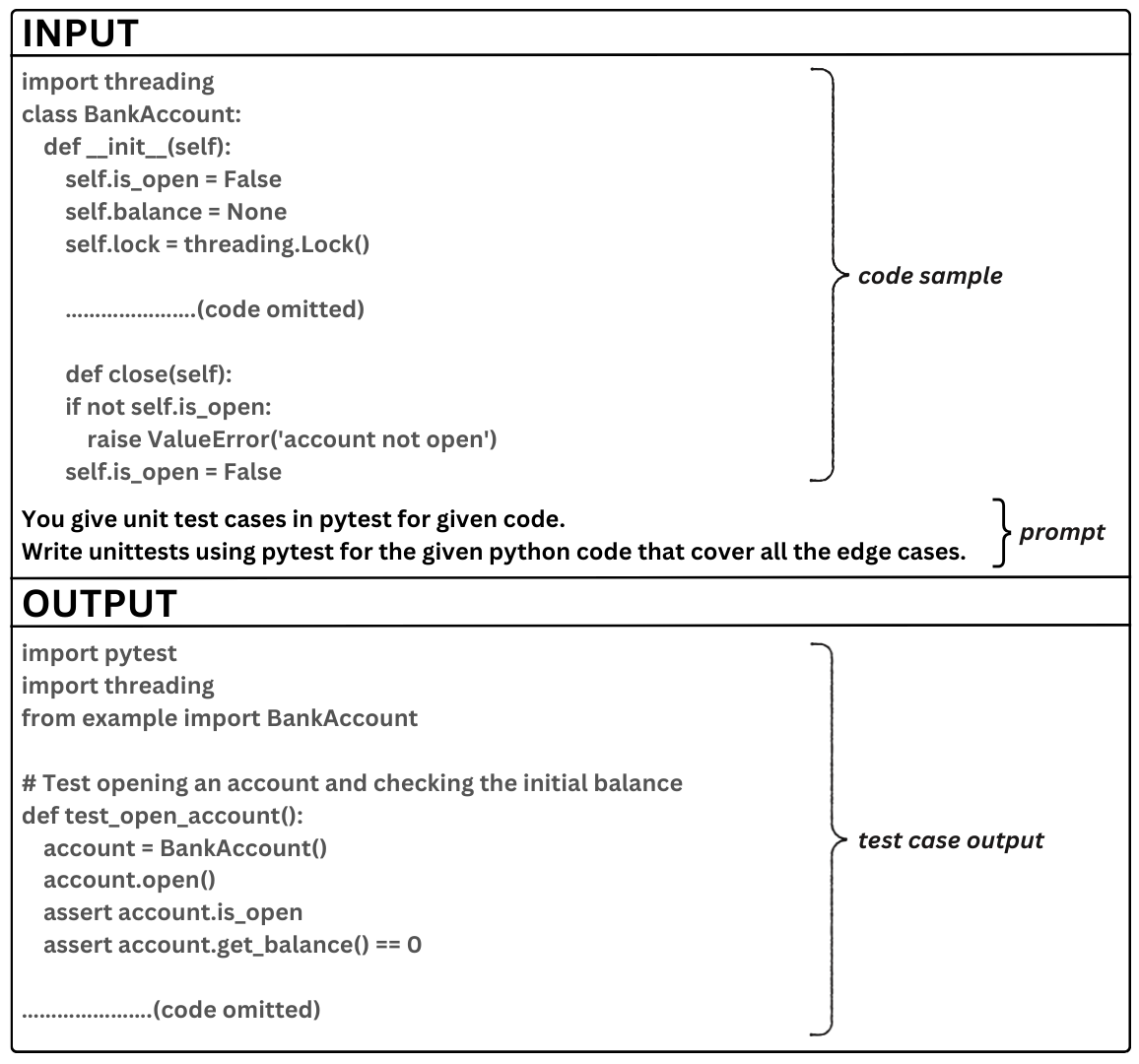}
  \caption{\textit{Basic Prompt}}
  \label{fig:basic-prompt}
\end{figure}

Our prompt comprises of two components: i) the Python program, and ii) the descriptive text in natural language outlining the task we aim to accomplish as shown in \textbf{Figure 1}.
In part i) we provide the whole code of the selected core module (100-300 LOC).  We are not providing separate units into ChatGPT, as many studies have\cite{Guilherme2023initialAnalysis}\cite{tang2023comp1}\cite{xie2023comp2}\cite{yuan2023comp3}, as we aim to evaluate ChatGPT’s ability to identify units when provided with a complete Python Program. In part ii), we query ChatGPT as follows: “Write Unit tests using Pytest for given Python code that covers all the edge cases."

The ChatGPT-generated test cases are then evaluated for their statement and branch coverage against that of Pynguin-generated test cases; through this, we also find the missed statements by ChatGPT and Pynguin, that the generated test cases are unable to cover, and see if they overlap. Next, we designed a new prompt for ChatGPT that would take in the indices of these missed statements, and ask it to again generate unit tests so as to improve the coverage. Following this, we are also piqued by the possibility of iteratively prompting ChatGPT to keep improving the coverage. We then repeatedly prompt ChatGPT while updating the prompt with the indices of the new set of missed statements after every iteration, till we observe no further improvement in coverage as illustrated in \textbf{Figure 2}.

\begin{figure}
  \centering
  \includegraphics[width=0.4\textwidth]{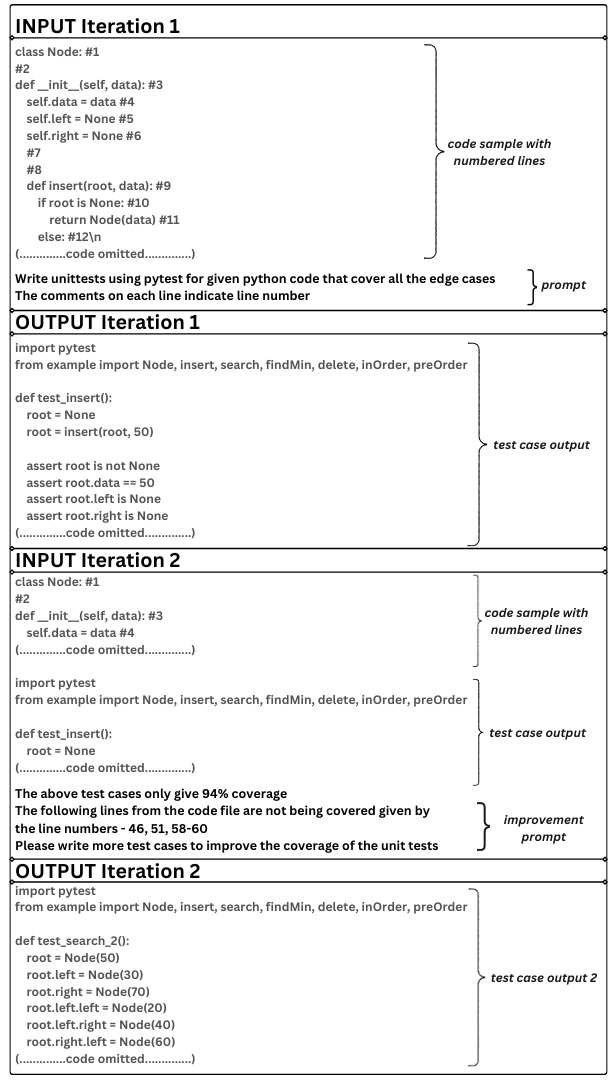}
  \caption{Improvement Prompt}
  \label{fig:improvement-prompt}
\end{figure}

\subsection{Evaluation Metrics}
The efficacy of the generated unit tests was assessed through a multifaceted approach, employing the following metrics:
\begin{itemize}
    \item \textbf{Statement Coverage:} Quantifying the extent to which the generated tests covered individual code statements.
    \item \textbf{Branch Coverage:} Evaluating the coverage of various code branches, and gauging the effectiveness of the test suite in exploring different execution paths.
    \item \textbf{Correctness:} Checking if the generated assertions are useful for evaluating the intended functionality of the code, in addition to being correct.
\end{itemize}

\subsection{Experimental Procedure}
After gathering the test cases generated by ChatGPT and Pynguin, we compare their performance based on statement and branch coverage. We then try to iteratively prompt ChatGPT to improve its coverage. After reaching the saturation point, where-after no improvement is observed in ChatGPT-generated tests, we evaluate the quality of the ChatGPT-generated test cases by looking at their correctness. We further find out whether any overlap exists in the missed statements between ChatGPT and Pynguin. \textbf{Figure 3} shows the workflow of our empirical analysis.

\begin{figure*}
  \centering
  \resizebox{\linewidth}{!}{
  \includegraphics[width=0.4\textwidth]{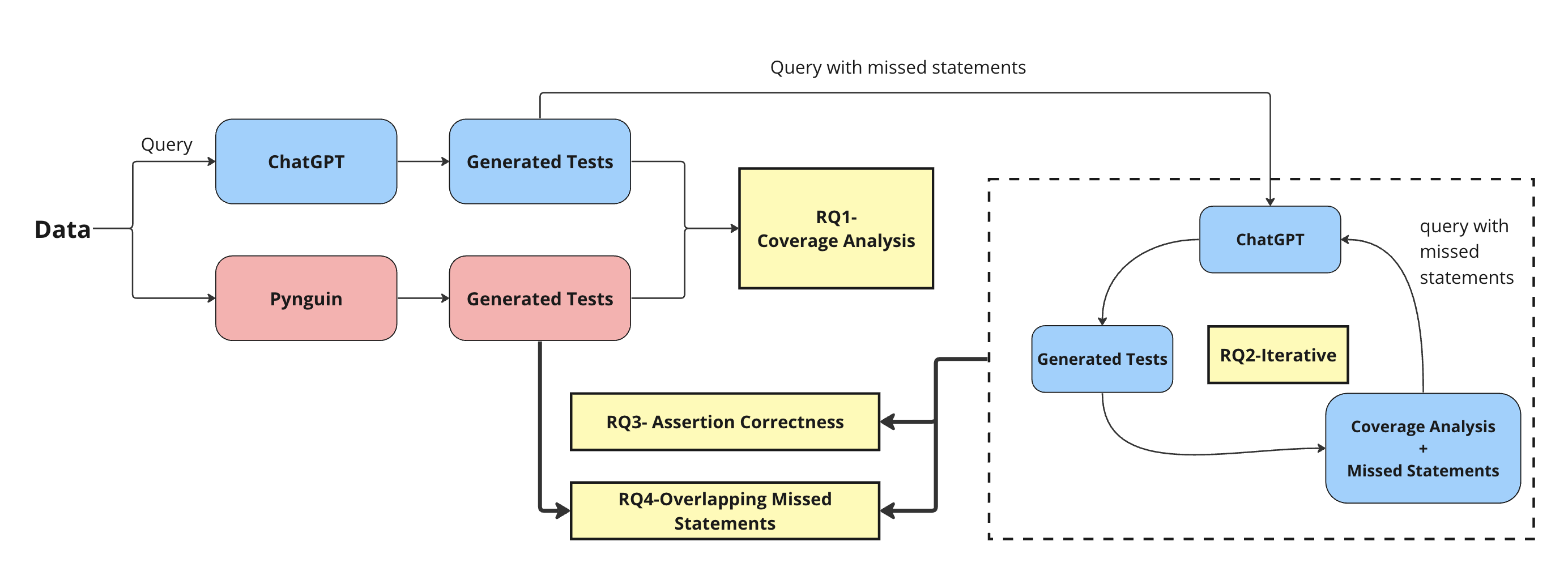}
  }
  \caption{\textit{Workflow of our Empirical Analysis}}
  \label{fig:flowchart}
\end{figure*}

Moving through, we aim to address the research questions in the following order:
\begin{itemize}
    \item \textbf{RQ1 (Comparative Performance):} We measure the differences in statement coverage and branch coverage between unit tests generated by ChatGPT and Pynguin across different code structures and complexities. We then investigate the correlation, if any, between cyclomatic complexity, code structure, and ChatGPT's achieved coverage.
    \item \textbf{RQ2 (Performance Saturation and Iterative Improvement):} We then iteratively prompt ChatGPT to improve the coverage for a sample, given the indices of missed statements.
    \item \textbf{RQ3 (Quality Assessment):} We check the generated unit test cases for compilation errors. And among the test cases that are compiling, we check whether their assertions correctly test the intended functionality of the code.
    \item \textbf{RQ4 (Combining Pynguin and ChatGPT for Improved Performance):} In cases where statements are missed, we look at the extent of overlap in missed statements between unit tests generated by ChatGPT and Pynguin. We use this to conclude whether a combination of techniques used by both tools could be a prospective solution for improved performance.
\end{itemize}

\section{Results}\label{sec:results}
\subsection{Small Code Samples (0-100 LOC)}
For Category 1 (procedural scripts), Pynguin failed to generate any unit tests cases. This can be attributed to the structure of Category 1 samples, where there is a lack of structure and Pynguin is unable to identify distinct units for testing as it relies on properties of modular code. It was noted that ChatGPT provides recommendations for refactoring the category 1 programs into modular units. It first generated the refactored code for the provided code sample and then proceeded to generate the test cases according to the modified code. 

For Category 2, there is no significant difference in the statement and branch coverage achieved by Pynguin and ChatGPT. This is also evident from the p-values (threshold = 0.05) obtained after performing Independent t-test \cite{t-test}. p-value for statement coverage is 0.631 while it is 0.807 for branch coverage. Both p-values are higher than the threshold.

For Category 3 also, there is no significant difference in statement and branch coverage achieved by the two tools, given the respective p-values are greater than the threshold: 0.218 and 0.193. 

In conclusion, ChatGPT and Pynguin give similar coverage for 0-100 LOC code samples as shown in \textbf{Table 1}.

\begin{table}[]
\centering
\resizebox{\columnwidth}{!}{%
\begin{tabular}{@{}ccccc@{}}
\toprule
\multirow{2}{*}{\textbf{0-100 LOC}}
& \multicolumn{2}{l}{\textbf{Avg Statement Coverage}} & \multicolumn{2}{l}{\textbf{Avg Branch Coverage}} \\
& \textbf{ChatGPT}     & \textbf{Pynguin}    & \textbf{ChatGPT}   & \textbf{Pynguin}   \\\midrule
\textbf{Category 1 (original)}   & 0     & 0    & 0     & 0     \\ \midrule
\textbf{Category 1 (refactored)} & 97.45 & 0    & 96.85 & 0     \\ \midrule
\textbf{Category 2}              & 93.26 & 90.3 & 91.68 & 90.1  \\ \midrule
\textbf{Category 3}              & 91.55 & 97   & 89.5  & 96.15 \\ \bottomrule
\end{tabular}%
}
\caption{\textit{Average statement and branch coverage obtained by ChatGPT \& Pynguin for small code samples. \textit{Both the tools give comparable performance for Category 2 and Category 3.}}}
\label{Table 4.1}
\end{table}

\subsection{ Large Code Samples (100-300 LOC)}
\subsubsection{\textbf{Category-wise Coverage Analysis}}  
Since for Category 1, unit test generation is not feasible due to lack of well defined units, and code-refactoring is a wide domain, we limit our coverage analysis to category 2 and 3 for large code samples having 100-300 LOC. For Category 2, we observed that there was no significant difference between the statement and branch coverage achieved by ChatGPT and Pynguin, with respective p-values greater than 0.05 (threshold); 0.169 and 0.195. For Category 3 as well, ChatGPT and Pynguin gave similar coverage, signified by p-values 0.677 and 0.580 for statement and branch coverage respectively. These observations are presented in detail in \textbf{Table 2} and \textbf{Figure 4}. 

\begin{table}[]
\centering
\renewcommand{\arraystretch}{1.4}
\resizebox{\columnwidth}{!}{%
\begin{tabular}{@{}ccccc@{}}
\toprule
\multirow{2}{*}{\textbf{100-300 LOC}}
& \multicolumn{2}{c}{\textbf{Avg Statement Coverage}} & \multicolumn{2}{c}{\textbf{Avg Branch Coverage}}  \\
& \textbf{ChatGPT}     & \textbf{Pynguin}    & \textbf{ChatGPT}   & \textbf{Pynguin}   \\ \midrule
\textbf{Category 2}                & 77.44 & 88.77 & 74.77 & 86.22 \\ \hline
\textbf{Category 3}                & 77.4  & 79.6  & 73    & 76.15 \\ \hline
\textbf{Average on all 40 samples} & 77.43 & 81.63 & 73.39 & 78.36 \\ \bottomrule
\end{tabular}%
}
\caption{\textit{Average statement and branch coverages obtained by ChatGPT \& Pynguin for large code samples. Here also, we find ChatGPT's coverage is comparable with Pynguin.}}
\label{}
\end{table}

\begin{figure*}
    \centering
    \begin{subfigure}{0.45\textwidth}
        \includegraphics[width=0.95\textwidth]{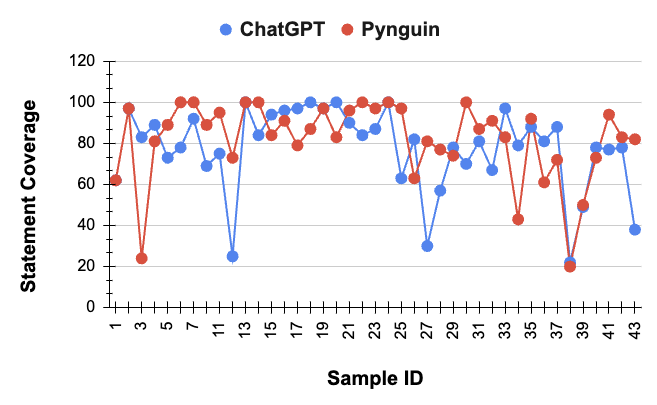}
        \caption{}
        \label{fig:statement-coverage-100-300}
    \end{subfigure}
    \begin{subfigure}{0.45\textwidth}
        \includegraphics[width=0.95\textwidth]{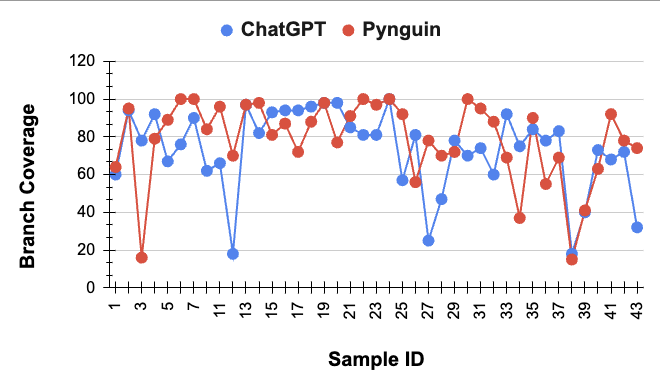}
        \caption{}
        \label{fig:branch-coverage-100-300}
    \end{subfigure}
    \caption{\textit{Statement coverage (left) and Branch coverage (right) obtained by ChatGPT (blue) and Pynguin (red) for all code samples (100-300 LOC).}}
\end{figure*}

To answer \textbf{RQ1}: ChatGPT and Pynguin give comparable statement and branch coverage for all 3 categories. Additionally, the Mccabe complexity is a metric to evaluate the complexity of a unit of code. For a code sample, we assign it the max Mccabe complexity, which is the maximum of all the units present in the code. However, plotting the average statement coverage against this complexity measure does not seem to highlight any trend or correlation between coverage achieved by each tool and the maximum Mccabe complexity of a code sample as seen in \textbf{Figure 5}. 

\begin{figure}
  \centering
  \includegraphics[width=0.4\textwidth]{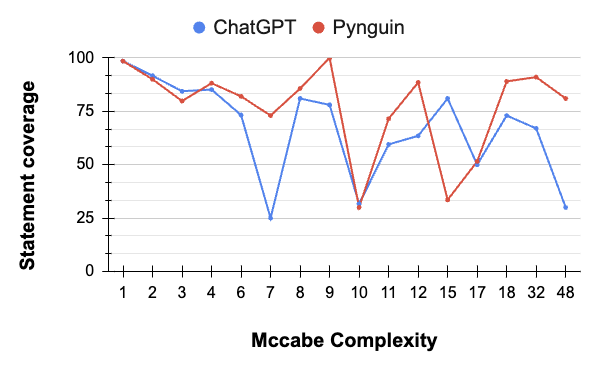}
  \caption{\textit{Statement coverage obtained by ChatGPT (blue) and Pynguin (red) for all code samples at different Mccabe Complexities.}}
  \label{fig:mccabe-vs-statement}
\end{figure}

\subsubsection{\textbf{Iterative improvement in Coverage}}
Till now, the coverage analysis was done on the results obtained from the first iteration of prompting ChatGPT. Providing the missed statements from the first iteration, as part of the prompt to ChatGPT, we ask it to further improve the coverage for a given code sample. We continue this process till there was no improvement in coverage between consecutive iterations. We had to iteratively prompt ChatGPT for 5 times at most since the coverage for most of the samples converged at iteration 4 as seen in \textbf{Figure 6}. 

To answer \textbf{RQ2:} It was observed that the statement coverage in Category 2 and Category 3 increased by 27.95 and 15.25 respectively on average. For Category 1, we found that despite the number of iterations there was no improvement in coverage at any step as seen in \textbf{Table 3}.

\begin{figure}
  \centering
  \includegraphics[width=0.4\textwidth]{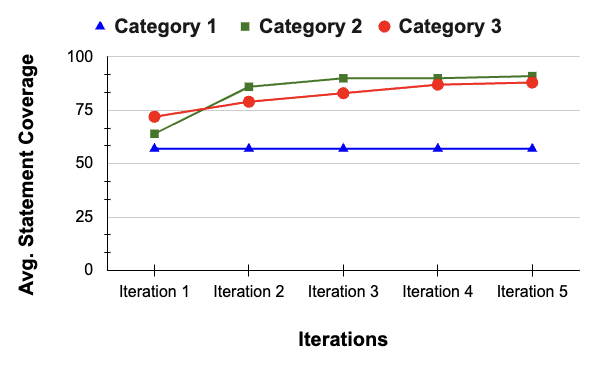}
  \caption{\textit{Average statement coverage obtained after each iteration for all of the 3 categories. We see that improvement in coverage saturates at 4 iterations.}}
  \label{fig:iterative-improvement}
\end{figure}

\begin{table}[]
\centering
\renewcommand{\arraystretch}{1.5}
\resizebox{\columnwidth}{!}{%
\begin{tabular}{cccccc}
\hline
\multirow{2}{*}{\textbf{Avg Coverage}} &
  \multirow{2}{*}{\textbf{Iter 1}} &
  \multirow{2}{*}{\textbf{Iter 5}} &
  \textbf{Best} &
  \textbf{Diff in best} &
  \textbf{Med iters for} \\
                                  &        &        & \textbf{Coverage} & \textbf{\& least cov} & \textbf{Cov plateau} \\ \hline
\textbf{Category 1 (100-300 LOC)}   & 56.833 & 56.833 & 56.833            & 0                    & 1                    \\ \hline
\textbf{Category 2 (100-300 LOC)} & 63.6   & 90.5   & 91.55             & 27.95                & 4                    \\ \hline
\textbf{Category 3 (100-300 LOC)} & 72.45  & 87.7   & 87.7              & 15.25                & 4                    \\ \hline
\end{tabular}%
}
\caption{\textit{Iterative Improvement. This table shows the Average statement coverages achieved in iteration 1 and 5 for all 3 categories. Also depicts the average improvement in statement coverage after 5 iterations and the median number of iterations it takes to reach the saturation point in coverage improvement. Best coverage: Average of Best Overall Coverage over all iterations. Diff in best \& least cov: Average Difference in best \& least coverage. Med iters for Cov plateau: Median iterations to achieve Coverage saturation.}}
\label{tab:my-table}
\end{table}

\subsubsection{\textbf{Correctness}}
To assess the correctness of assertions produced by ChatGPT, we examine the various error categories in the following manner: i) whether the test cases are compiling, ii) the percentage of passing assertions among the compiling test cases, and iii) the nature of errors encountered for assertions that fail. It is crucial to emphasize that the source code snippets used for generating these assertions are derived from well-established Python projects publicly available for general use, meaning that these projects must have been thoroughly tested to perform what they were intended to do. This implies that the generated test cases by ChatGPT should effectively capture the intended functionality of the code, and the corresponding assertions should pass if they correctly test the logic of the code. However, if any assertions do fail, it indicates that those assertions fail to test the intended functionality of the code and thus are not correct. 

\begin{table}[]
\centering
\renewcommand{\arraystretch}{1.4}
\resizebox{\columnwidth}{!}{%
\begin{tabular}{cccc}
\hline
\textbf{ChatGPT Generated}         & \textbf{Category 1}  & \textbf{Category 2}    & \textbf{Category 3}    \\
\textbf{\% of incorrect assertions}  &    &    &    \\ \hline
\textbf{All Incorrect}    & 57.75 & 39.4  & 27.67 \\ \hline
\textbf{Assertion Error}  & 86.93 & 78.30 & 70.96 \\ \hline
\textbf{Try/Except Error} & 0     & 2.68  & 3.66  \\ \hline
\textbf{Runtime error}    & 13.07 & 19.02 & 25.38 \\ \hline
\end{tabular}%
}
\caption{\textit{Percentage of Incorrect Assertions. This table gives the average percentage of incorrect assertions generated by ChatGPT for all 3 categories of code samples. Also specifies the distribution across 3 causes of failing assertions: i) AssertionError: occurs when the asserted condition is not met ii) Error in try/except block: occurs when an exception was expected to be raised, but it wasnt raised iii) Runtime Error : occurs while the program is running after being successfully compiled.}}
\label{tab:my-table}
\end{table}

To answer \textbf{RQ3}: As depicted in \textbf{Table 4}, about 39\% of generated assertions are incorrect on average for Category 2 while 28\% of assertions are incorrect on average for Category 3. Separately, we also checked the correctness of assertions for Category 1 samples, for which ChatGPT had provided refactored code, and found that 58\% of the generated assertions were incorrect. The decrease in percentage of incorrect assertions as we go from category 1 to category 3, may imply that ChatGPT’s ability to generate correct assertions is higher for programs with well defined structure, possibly due to presence of more coherent and meaningful units in the code.

\subsubsection{\textbf{Overlapping Missed Statements}}
We looked at the number of overlapping statements that were commonly being missed by ChatGPT and Pynguin. We observed, on average, out of a combined total of 50 missed statements, around 17 were common to both, which means the overlap is significantly lower than the total number of missed statements. 

To answer \textbf{RQ4:} If we were to combine ChatGPT and Pynguin, a possible logical inference would be that, the only missed statements would be the ones which were overlapping, and rest all the statements, which were earlier being individually missed by ChatGPT and Pynguin, will now be covered. For example, based on our evaluation, a combination of the two tools will result in 31 more statements being covered.
This is illustrated in \textbf{Table 5} and \textbf{Figure 7}. This is also in line with the findings of CODAMOSA \cite{lemieux2023codamosa} which proposes that a combination of SBST and LLM can lead to better coverage.

\begin{table}
\centering
\begin{tabularx}{\columnwidth}{Xc}
\toprule
\textbf{Intersecting missed} & \textbf{17.78} \\ \midrule
\textbf{Union missed}     & \textbf{50.33} \\ \midrule
\textbf{Minimum missed}    & \textbf{20.29} \\ \midrule
\textbf{Minimum covered}   & \textbf{2.51}  \\ \midrule
\textbf{Maximum covered}   & \textbf{31.64}  \\ \bottomrule
\end{tabularx}
\caption{\textit{Summary Statistics of missed overlapping statements. This table shows summary statistics on statements not covered through the tests generated by ChatGPT and Pynguin. Intersecting missed: average number of common statements that were missed by both ChatGPT and Pynguin. Union missed: average number of statements that were missed by either ChatGPT or Pynguin. Minimum missed: Min of (average number of statements missed by ChatGPT, average number of statments missed by Pynguin). Minimum covered: The number of statements that are atleast covered after combining ChatGPT and Pynguin on average. Maximum covered: It is the maximum possible number of statements that should be covered after combining Chatgpt and Pynguin on average.}}
\label{tab:my-table}
\end{table}


\begin{figure}
  \centering
  \includegraphics[width=0.4\textwidth]{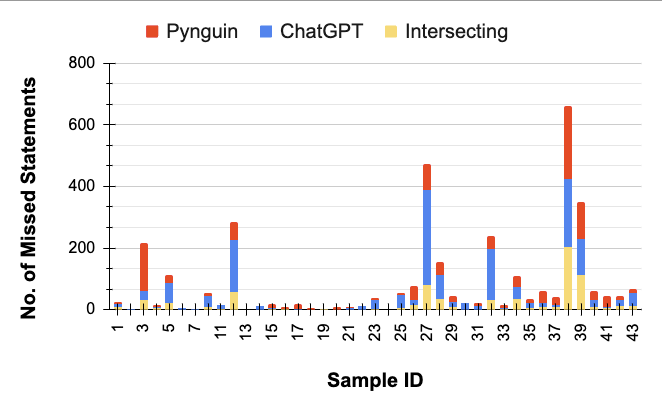}
  \caption{\textit{Missed Statements by ChatGPT, Pynguin and their intersection for each code sample.}}
  \label{fig:missed-statements}
\end{figure}


\section{Related Work}\label{sec:rw}
In this section, we briefly discuss some of the existing tools and techniques for generating unit tests. 

\noindent \textbf{Search-based software testing (SBST) techniques:} These techniques \cite{Fraser2012SBST3, Andrews2011SBST2, Baresi2010SBST4, derakhshanfar2022SBST1, Tonella2004SBST5} turn testing into an optimization problem to generate unit test cases. The goal of SBST is to generate optimal test suites that improve code coverage and efficiently reveal program errors by utilizing algorithms to traverse problem space. This approach shows potential in lowering the quantity of test cases needed while preserving reliable detection of errors. EvoSuite \cite{Fraser2011Evosuite} is an automated test generation tool which utilizes SBST. It takes a Java class or method as input, uses search-based algorithms to create a test suite meeting specified criteria (e.g., code or branch coverage), and evaluates test fitness. Through iterative processes of variation, selection, and optimization, EvoSuite generates JUnit test cases and provides a report on the effectiveness of the produced test suite based on metrics like code coverage and mutation score.

Pynguin \cite{Lukasczyk_2022Pynguin} is another tool which utilizes SBST for generating unit tests in Python programming language. The variable types are dynamically assigned at runtime in Python which makes it difficult to generate unit tests. Pynguin examines a Python module to gather details about the declared classes, functions, and methods. It then creates a test cluster with every relevant information about the module being tested, and during the generation process, chooses classes, methods, and functions from the test cluster to build the test cases. We use Pynguin as our baseline for comparing the suitability and effectiveness of LLMs in unit test generation.


\noindent \textbf{Randomized test generation techniques:} Randoop \cite{Pacheco2007Randoop} uses feedback-directed random testing to generate test cases. The basic idea behind this technique is to generate random sequences of method calls and inputs that exercise different paths through the program. As the test runs, Randoop collects information about the code coverage achieved by the test, as well as any exceptions that are thrown. Based on this feedback, Randoop tries to generate more test cases that are likely to increase code coverage or trigger previously unexplored behaviour.

\noindent\textbf{AI-based techniques:} Ticoder \cite{lahiri2023ticoder} presents an innovative Test-Driven User-Intent Formalisation (TDUIF) approach for generating code from natural language with minimal formal semantics. Their system, TICODER, demonstrates improved code generation accuracy and the ability to create non-trivial functional unit tests aligned with user intent through minimal user queries. TOGA \cite{Dinella_2022toga} introduces a neural method for Test Oracle Generation, using a transformer-based approach to infer exceptional and assertion test oracles based on the context of the focal method. 

CODAMOSA \cite{lemieux2023codamosa} introduces an algorithm that enhances Search-Based Software Testing (SBST) by utilizing pre-trained large language models (LLMs) like OpenAI's Codex \cite{Codex}. The approach combines test case generation with mutation to produce high-coverage test cases and requests Codex to provide sample test cases for under-covered functions. Our paper confirms some of the findings from CODAMOSA. For instance, our results also show that a combination of LLM and Pynguin (SBST-based) can lead to a better coverage. At the same time, CODAMOSA does not explore some of the research questions which we explored in this paper such as (1) How correct are the assertions generated by LLMs? (2) Do the LLM-generated tests align with the intended functionality of the code? (3) How does the performance of LLM improve over multiple iterations of prompting? 

\section{Conclusion}\label{sec:conclusion}
In this study, we discovered that ChatGPT and Pynguin demonstrated nearly identical coverage for both small and large code samples, with no statistically significant differences in average coverages across all categories. When iteratively prompting ChatGPT to enhance coverage, by providing the indices of missed statements from previous iteration, improvements were notable for categories 2 and 3, reaching saturation at 4 iterations, while no improvement occurred for category 1. 

Notably, individually missed statements by both tools showed minimal overlap, hinting at the potential for a combined approach to yield higher coverage. Lastly, our assessment of the correctness of ChatGPT-generated tests revealed a decreasing trend in the percentage of incorrect assertions from Category 1 to 3, which could possibly suggest that assertions generated by ChatGPT are more effective in cases where code units are well defined. 

ChatGPT operates with a focus on understanding and generating content in natural language rather than being explicitly tailored for programming languages. While ChatGPT may be capable of achieving high statement coverage in the generated unit tests, a high percentage of the assertions within those tests might be incorrect. A more effective approach to generating correct assertions would be based on the actual semantics of the code. This presents a concern that ChatGPT may prioritize coverage over the accuracy of the generated assertions, which is a potential limitation in using ChatGPT for generating unit tests, and a more semantic-based approach might be needed for generating accurate assertions. Future research endeavors could delve into several promising avenues based on the findings of this study. Firstly, exploring how ChatGPT refactors code from procedural scripts and assessing whether the refactored code preserves the original functionality could provide valuable insights into the model's code transformation capabilities. Additionally, investigating the scalability of ChatGPT and Pynguin to larger codebases and more complex projects may offer a broader understanding of their performance in real-world scenarios. Furthermore, a comprehensive exploration of the combined use of ChatGPT and Pynguin, considering their complementary strengths, could be undertaken to maximize test coverage and effectiveness. Lastly, examining the generalizability of our observations across diverse programming languages and application domains would contribute to a more comprehensive understanding of the applicability and limitations of these tools.

\bibliographystyle{ACM-Reference-Format}
\bibliography{references}


\end{document}